\documentclass{aa} 

\usepackage{lscape}
\usepackage{graphicx}
\usepackage{txfonts}
\DeclareUnicodeCharacter{00A0}{ }

\begin{document}

   \title{The injection of ten electron/$^{3}$He-rich SEP events}

   \author{Linghua Wang\inst{1}
           \and S\"am Krucker\inst{2,3}
           \and Glenn M. Mason\inst{4}
           \and Robert P. Lin\inst{2}
           \and Gang Li\inst{5}
          }

   \institute{Institute of Space Physics and Applied Technology, Peking University, Beijing, 100871, China\\
              \email{wanglhwang@gmail.com}
         \and
          Space Sciences Laboratory, University of California, Berkeley, CA 94720, USA
         \and    
          Institute of 4D Technologies, University of Applied Sciences Northwestern Switzerland, 5210 Windisch, Switzerland
         \and
          Applied Physics Laboratory, Johns Hopkins University, Laurel, MD 20723, USA
         \and
          Department of Space Science and CSPAR, University of Alabama in Huntsville, Huntsville, AL 35899, USA
             }

   \date{}
 
  \abstract{We have derived the particle injections at the Sun for ten good electron/$^{3}$He-rich solar energetic particle (SEP) events, using a 1.2 AU particle path length (suggested by analysis of  the velocity dispersion). The inferred solar injections of high-energy ($\sim$10 to 300 keV) electrons and of $\sim$MeV/nucleon ions (carbon and heavier) start with a delay of 17$\pm$3 minutes and 75$\pm$14 minutes, respectively, after the injection of low-energy ($\sim$0.4 to 9 keV) electrons. The injection duration (averaged over energy) ranges from $\sim$200 to 550 minutes for ions, from $\sim$90 to 160 minutes for low-energy electrons, and from $\sim$10 to 30 minutes for high-energy electrons. Most of the selected events have no reported H$\alpha$ flares or GOES SXR bursts, but all have type III radio bursts that typically start after the onset of a low-energy electron injection. All nine events with SOHO/LASCO coverage have a relatively fast ($>$570km/s), mostly narrow ($\lesssim$30$^{\circ}$), west-limb coronal mass ejection (CME) that launches near the start of the low-energy electron injection, and reaches an average altitude of $\sim$1.0 and 4.7 $R_{S}$, respectively, at the start of the high-energy electron injection and of the ion injection. The electron energy spectra show a continuous power law extending across the transition from low to high energies, suggesting that the low-energy electron injection may provide seed electrons for the delayed high-energy electron acceleration. The delayed ion injections and high ionization states may suggest an ion acceleration along the lower altitude flanks, rather than at the nose of the CMEs.}
  


  \maketitle

%

\section{Introduction}

Impulsive Solar Energetic Particle (SEP) events (also called electron/$^{3}$He-rich SEP events) are dominated by $\sim$1-100 keV electrons and low-intensity $\sim$MeV/nucleon ion emissions with large enhancements of $^{3}$He ($^{3}$He/$^{4}$He up to $\sim$10$^{4}$ times the coronal values), heavy nuclei, such as Fe (factor of $\sim$10) and ultra-heavy nuclei up to $>$200 amu (factor of $>$200), and high ionization states (e.g., Fe$^{20+}$) \citep[see][for review]{Mas07}. At solar maximum, $\gtrsim$150 solar electron events/year are observed near the Earth with a longitude extent of $\sim$30$^{\circ}$-60$^{\circ}$ \citep{Wang12}, implying the occurrence of $\sim$10$^{4}$ events/year over the whole Sun. This makes electron/$^{3}$He-rich SEP events the most common solar particle acceleration phenomenon. 

\citet{Ream95, Ream99} suggested that impulsive SEP events are produced in impulsive flares. However, \citet{Wang12} find that only about one third of $^{3}$He-rich electron events are associated with a reported GOES soft X-ray (SXR) flare, while essentially all ($\sim$99\%) are accompanied by type III radio bursts and $\sim$60\% are associated with west-limb coronal mass ejections (CMEs). Many studies also show that some impulsive SEP events are associated with coronal jets and narrow CMEs that originate from western hemisphere flaring active regions, which are next to coronal holes containing the Earth-connected open field lines \citep[e.g.,][]{Kah01, Wang06a, Pic06, Nit08}. These studies suggest that magnetic reconnection between closed and open field lines may be involved in producing electron/$^{3}$He-rich SEP events.

The timing of particle injection in impulsive SEP events carries important information for understanding the acceleration of electrons and $^{3}$He-rich ions. \citet{Krk99} and \citet{Hag02} found that the injection of the $>$25-38 keV electrons at the Sun was delayed by $\sim$10 to 30 minutes after the release of type III radio bursts in most ($\sim$80\%) of impulsive events. \citet{Wang06b} reported that the observed in situ flux-time profiles for three strongly scatter-free electron events fit well to an isosceles triangle injection profile (with equal rise and fall time) at the Sun, with the low-energy ($\sim$0.4-10 keV) electron injection starting $\sim$9 minutes before the type III bursts but the high-energy ($\sim$10-300 keV) electron injection starting $\sim$8 minutes after, suggesting that low-energy electrons generate the type III radio emissions. Moreover, \citet{Ream85} showed in one electron/$^{3}$He-rich SEP event that $^{3}$He-rich ions were injected close to the electron injection (within $\sim$1 hour), while \citet{Ho03} reported five electron/$^{3}$He-rich events with a delayed ($>$40 minutes) ion injection. 

In this paper, we survey the electron observations from the WIND 3-D Plasma and Energetic Particle (3DP) instrument and the ion observations from the ACE Ultra Low Energy Isotope Spectrometer (ULEIS) from November 1997 to December 2003, and select ten good electron/$^{3}$He-rich SEP events. For these events, we estimate the particle path length in the interplanetary medium (IPM) and derive the injection of electrons and $^{3}$He-rich ions at the Sun (Section 2). We also examine their association with H$\alpha$ flares, GOES SXR flares, CMEs and solar radio bursts (Section 3).

\begin{figure*}
\centering
\includegraphics[scale=0.95]{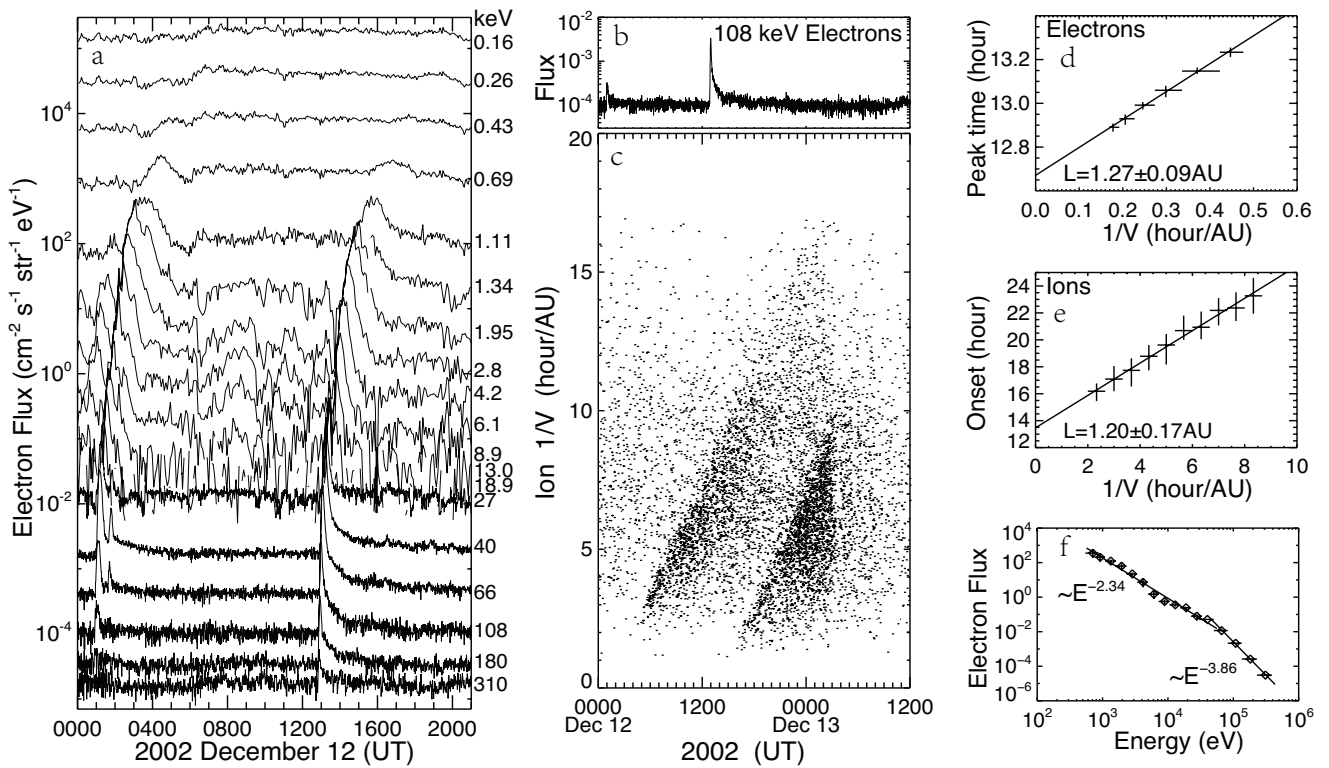}
\caption{Overview plot for event \#9 and 10. (a): The flux vs. time of electrons travelling outward from the Sun observed by WIND/3DP on December 12, 2002. The labels indicate the channel center energies. (b): The flux-time profile of the outward traveling electrons at 108 keV. (c): The ion spectrogram of 1/speed vs. arrival time for 10-70 amu ions measured by ACE/ULEIS on December 12-13, 2002. (d): The velocity dispersion analysis of the observed times of the electron peak flux at energies above 25 keV for event \#10. (e): The same analysis, but for the onset times of carbon and heavier ions. (f): The electron energy spectrum of background-subtracted peak flux for event \#10. In (d-e), the straight lines represent the least-squares fit to the data points. \label{fig1}}
\end{figure*}

\section{Observations}

The electron observations of a few eV to $\sim$400 keV used here are  from the 3DP instrument \citep{Lin95} - silicon semiconductor telescopes (above $\sim$25 keV) and electron electrostatic analyzers (below $\sim$25 keV) - on the WIND spacecraft. The ion observations of the elements He-Ni from $\sim$20 keV/nucleon to 10 MeV/nucleon are from the ULEIS instrument \citep{Mas98} on the ACE spacecraft. Both WIND and ACE spacecrafts are located in  solar wind near the Earth. In this paper, we utilize the measurements of carbon and heavier ions to achieve a better ion timing, since they generally have a lower background than the measurements of helium.

An ion event is defined to be associated with an electron event, if the in situ onset of the $\sim$2 MeV/nucleon ions ($V\sim0.065c$) occurs within four hours of the in situ onset of the $\sim$1 keV electrons ($V\sim0.062c$). We search through the electron and ion data from November 1997 through December 2003 (after 2003, the 27 keV electron channel begins to have significant  background noise), and find ten good electron/$^{3}$He-rich SEP events (Table~\ref{tbl-1}) that have: (1) a clear velocity dispersion (i.e., faster particles arriving at 1 AU before slower ones) for both electrons and ions, (2) a solar electron event detected from $\sim$0.4 keV to $\gtrsim$100 keV, and (3) a $^{3}$He/$^{4}$He ratio $>$0.1 at $\sim$0.5-2.0 MeV/nucleon.

Figure~\ref{fig1} shows two representative electron/$^{3}$He-rich SEP events (\#9 and 10) observed on December 12, 2002. The electron velocity dispersion is clearly evident from $\sim$0.4 keV (0.3 keV) up to 180 keV (310 keV) for event \#9 (\#10) measured by WIND/3DP (Figure~\ref{fig1}(a)). At all these energies, the observed electron temporal profiles exhibit nearly symmetric rapid-rise, rapid-decay peaks, with pitch-angle distributions strongly peaked outward along the interplanetary magnetic field (not shown), followed by slow-decay tails at flux levels that are much lower than the peak values. Such temporal profiles imply that, in these events, most of the electrons (those in the peak) propagate through the IPM essentially scatter-free \citep{Lin74}, and that the electron injection at the Sun would be impulsive and nearly symmetric \citep{Wang06b}. We note that the electron event \#9 is preceded by one smaller event that arrived about 25 minutes earlier at $\sim$1.3 - 6 keV, and is followed by another smaller event that started about 40 minutes later at $\sim$1.3 - 66 keV. 

Figure~\ref{fig1}(c) plots the spectrogram of the ion arrival time versus inverse speed for all the carbon and heavier ions measured by ACE/ULEIS on December 12-13, 2002. For events \#9 and 10, the ion velocity dispersion is also clearly evident, indicated by the leading edge of the observed ion arrivals at all energies. A sharp intensity-drop occurs around 0300 UT on December 13, 2002 simultaneously at all energies, likely due to the passage of an emptier flux rope with smaller particles intensities \citep{Maz00}. 

\begin{figure*}
\centering
\includegraphics[scale=0.95]{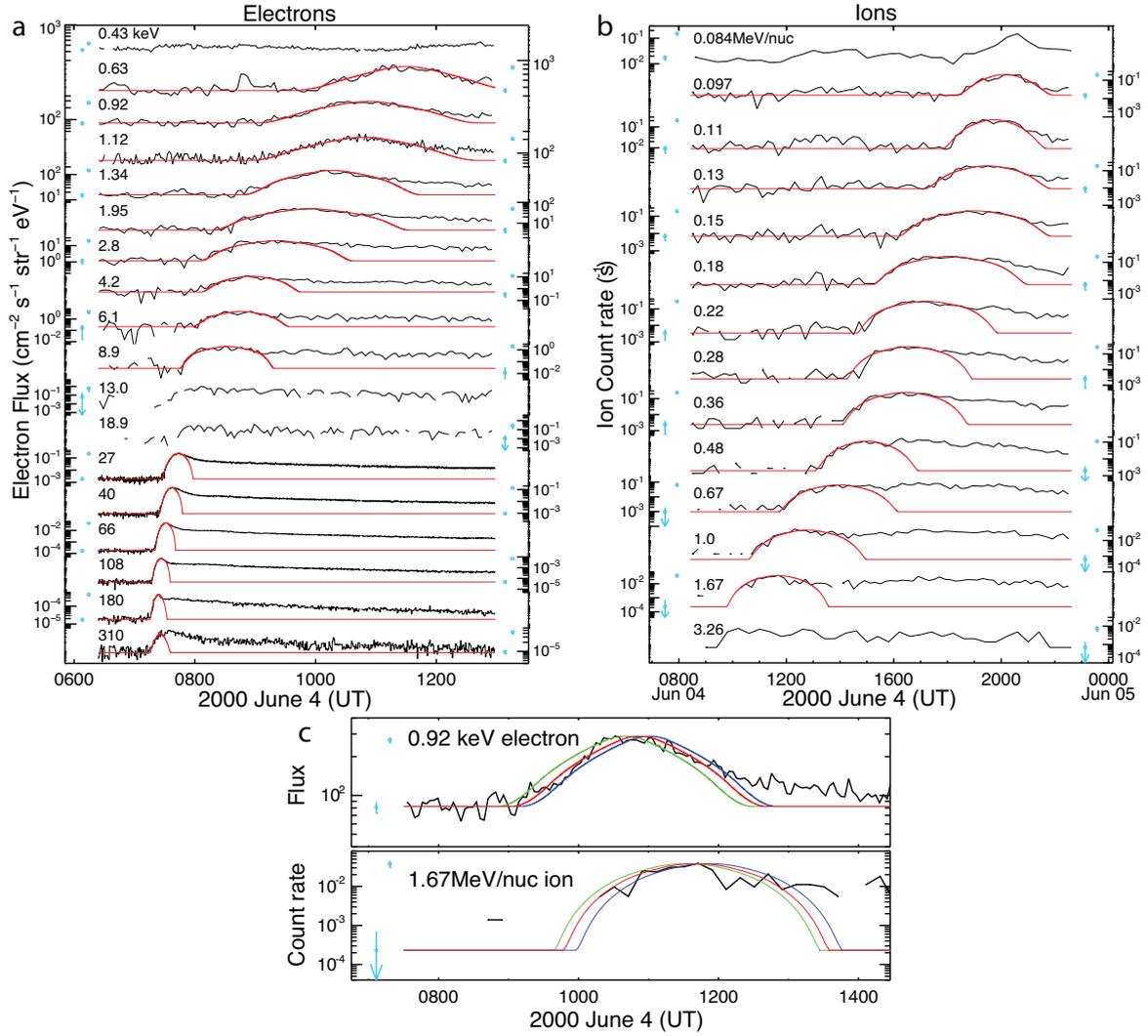}
\caption{Triangular fitting to the rapid-rise, rapid-decay peak of the temporal profiles of electron fluxes (a) and ion (carbon and heavier) count rates (b) observed at 1 AU, for event \#4. The temporal profiles at different energies have been shifted to achieve good separations. The black curves show the observations and the red curves are the best fits. The blue dots with uncertainties on the left or right of the panels represent the typical intensity of the pre-event background and of the peak with statistical errors. (c): The error analysis for the 0.92 keV electron channel and the 1.67 MeV/nucleon ion channel. The green and blue curves are the upper and lower limits of fitting (that represent a 95\% confidence interval), respectively. The limits on the start time of the corresponding injection are shown as error bars in Figure~\ref{fig3}(a).\label{fig2}}
\end{figure*}

The particle path length traveled from the Sun to the spacecraft, $L$, can be estimated from the velocity dispersion, i.e., $L=V_{i}[t_{i}-t_{0}]$, where $t_0$ is the estimated timing of particle injection at the Sun, $V_i$ is the particle velocity and $t_i$ is the observed timing at 1 AU for the $i$th energy channel. For these selected electron events, the peak to background intensity ratio is generally $>$ 10-100 at energies above 25 keV (see Figure 2(a) for example); at $>$ 25 keV, we use the observed peak-flux times at 1 AU as $t_{i}$ since the determination of electron peaks is more precise than the onset determination (because of the varying background of interplanetary superhalo electrons with energy \citep{Wan15}). For these ion events, the peak-to-background intensity ratio (carbon and heavier) is generally $>$ 10 (see Figure 2(b) for example); we use the observed onset times as $t_{i}$ since the onset can be determined more precisely for $\sim$MeV/nucleon heavy ions. For event \#10, the derived $L$ is 1.27$\pm$0.09 AU for electrons and 1.20$\pm$0.17 AU for ions (Figure 1(d) and (e)), consistent with the 1.19 AU smooth spiral field length for the observed solar wind speed of 370 km/s. For the ten selected  electron/$^{3}$He-rich events, the velocity dispersion analysis gives similar path lengths, consistent with a nominal 1.2 AU length for the Parker spiral within the estimated uncertainties ($\sim$0.1-0.2 AU). Hence, in this study we use a $L$ of 1.2 AU. i.e., the nearly scatter-free transport, for all ten events.

For nearly scatter-free SEP events with good statistics, we can obtain the particle injection profiles at the Sun from fitting to the rapid-rise, rapid-fall peak of in situ observations at 1 AU \citep{Wang06b}. For energy channel $j$, we assume an isosceles triangle-shaped injection profile (with equal rise and fall times) at the Sun:
\begin{eqnarray}
f_j&=&
\begin{cases} 
A_j\cdot E^{-\beta}\cdot\left(1-\dfrac{|t-t_{0j}-\Delta t_j/2|}{\Delta t_j/2}\right) &\text{ 0$\leq$$t-t_{0j}$$\leq$$\Delta$$t_j$;}\nonumber\\ 
0 &\text{ otherwise,}  \nonumber
\end{cases}
\end{eqnarray}
where $t_{0j}$ and $\Delta t_j$ are the start time and time duration of particle injection; $A_j\cdot E^{-\beta}$ represents the injection flux (or count rate) intensity, and $\beta$ is the index of particle peak flux (or count rate) spectrum observed at 1 AU. After taking into account the response to the energy bandwidth, the particle intensity $F_{j}$ at the propagation distance $L$ from the Sun is calculated as
\begin{equation}
\label{eq:prf}
F_{j}(L,t)= \frac{\int_{E_{j-}}^{E_{j+}} f_{j} \left(E,t_{0j}+\frac{L}{v(E)},\Delta t_{j},A_{j},t \right) dE}{E_{j+}-E_{j-}}.\nonumber
\end{equation}
The solar injection profile, $f_j$, is determined by best fitting $F_{j}(1.2AU,t)$ to the observed particle intensity-time profile at 1 AU, varying the start time $t_{0j}$, duration $\Delta t_{j}$, and intensity $A_{j}$. Hereafter this forward-fitting analysis is referred to as "triangular fitting".

\begin{figure*}
\centering
\includegraphics[scale=0.95]{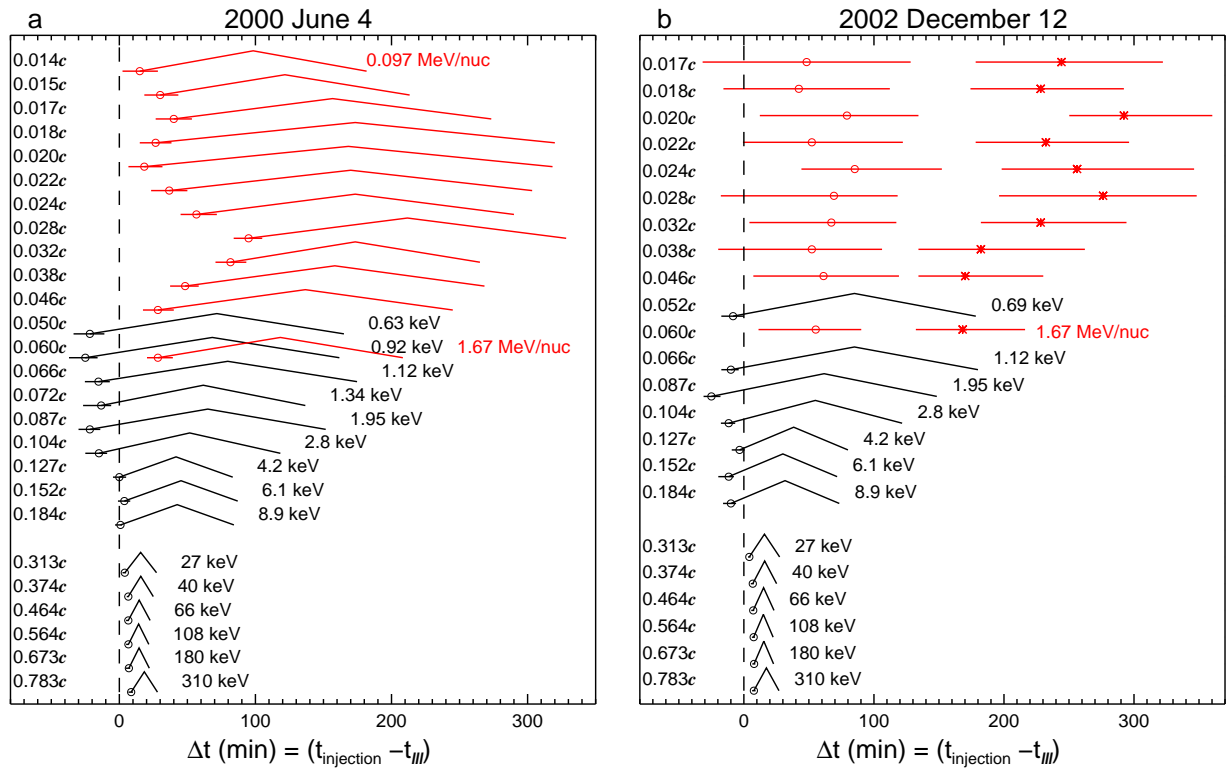}
\caption{Inferred solar injection timing of electrons (black) and ions (carbon and heavier, red) for event \#4 (a) and \#10 (b). The x axis shows the time in minutes, with respect to the estimated coronal release time of type III radio bursts (dashed line at $\Delta t$ = 0). The triangular profiles indicate the time variation of particle injections in intensity. The circles indicate the start time of particle injections. In panel (b), the asterisks represent the peak time of ion injections. The labels indicate the energy or the velocity in terms of speed of light $c$, for different energy channels. \label{fig3}}
\end{figure*}

\begin{figure}
\includegraphics[scale=0.8]{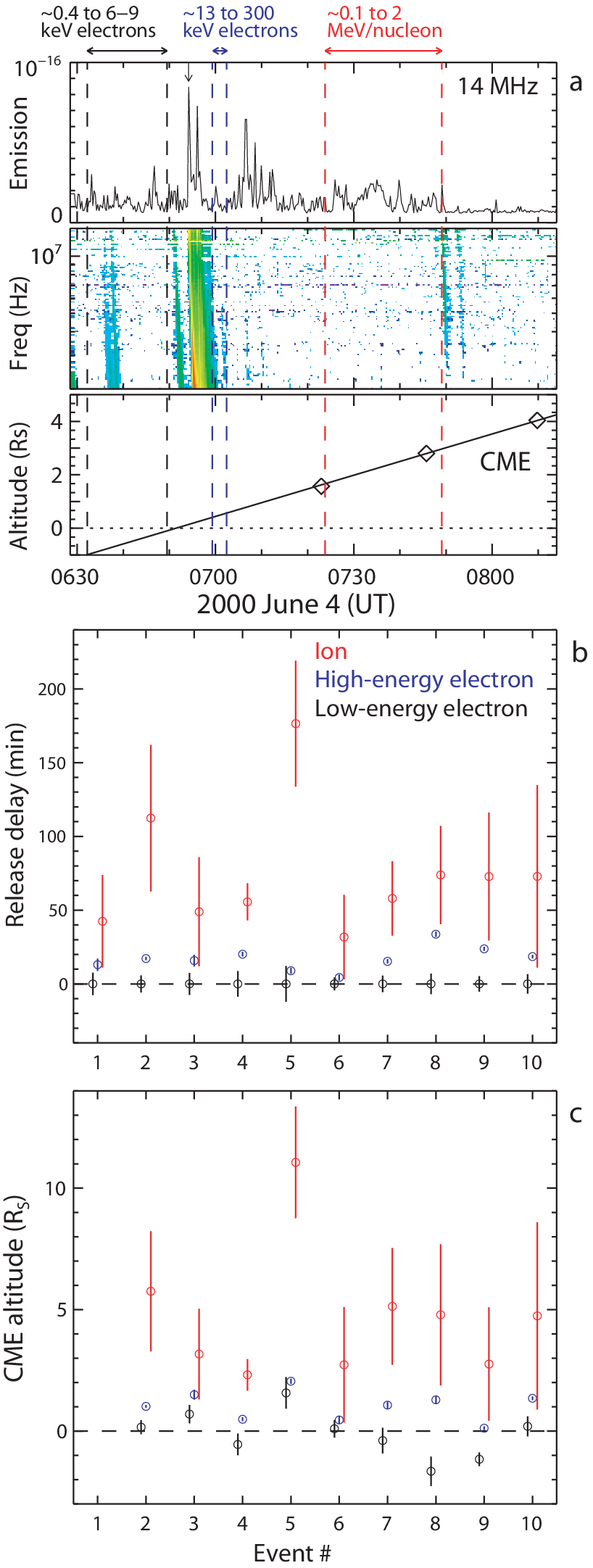}
\caption{Panel (a): Temporal comparison for event \#4. Top: The radio emission intensity measured at $\sim$14 MHz by WIND/WAVES, with the black vertical arrow indicating the time of maximum emission. Middle: The dynamic radio spectrogram at decametric to hectometric wavelengths. Bottom: The CME altitude above the photosphere measured by SOHO/LASCO. The  radio and CME data shown have been shifted back by the photon travel time of 500 s from the Sun to Earth.  Panel (b): Particle injection delays after the start of low-energy  electron injection for the ten events. Panel (c): Projected CME altitudes above the photosphere (horizontal dashed line) at the start of particle injections. Vertical dashed lines in (a) and circles (with error bars) in (b) and (c) indicate the start-time range of inferred solar injections for low-energy electrons (black), high-energy electrons (blue) and carbon and heavier ions (red).\label{fig4}}
\end{figure}

For the June 4, 2000 event (\#4), this triangular fitting is applied to electron energy channels from 0.63 to 310 keV (Figure~\ref{fig2}(a)), excluding the 13 and 18.9 keV channels because of poor statistics at those energies, and to ion channels from $\sim$0.1 to 1.67 MeV/nucleon (Figure~\ref{fig2}(b)). The fits (red curves) are good through the rapid-rise, rapid-fall phase for both electrons and ions. Afterwards, the observed electron fluxes show a strong, much slower decay at energies above $\sim$2 keV, mainly because of electrons reflected/scattered between the Earth's bow shock \citep{Kat13} and the converging interplanetary magnetic field within 1 AU. At 310 keV (probably also at 180 keV), a second electron peak is detected about ten minutes after the first peak (clearly shown in the linear flux scale), possibly as a result of a second solar electron injection that has a much harder spectrum, and is thus not detectable during the slow decay of the first peak at lower energies; these injections are associated with multiple type III bursts shown in Figure~\ref{fig4}(a). At these energies, the electron triangular fitting is applied only to the first peak. For ions, the observed count rates show an increase that occurrs around 1600UT simultaneously at energies of 0.48-1.67 MeV/nucleon, probably caused by the passage of a flux rope with higher fluxes at those energies \citep{Maz00}. At these ion energies, the triangular fitting is applied only to the observed ion time profile before this increase.

Figure ~\ref{fig3}(a) plots the corresponding triangular injection profiles of electrons (black) and ions (red) at the Sun for event \#4. The uncertainties in the start time (open circles) are estimated from the upper and lower limits of fitting to the entire rise (from onset to peak) of in situ observations (Figure~\ref{fig2}c), including the effect of pre-event background. For event \#4, low-energy (below $\sim$10 keV) electron injections start 13.4$\pm$8.7 minutes before the coronal release, $t_{III}$, of type III radio burst (dashed line in Figure~\ref{fig3}a) and last for $\sim$80-190 minutes, while high-energy (above $\sim$10 keV) electron injections have clearly delayed starts (6.7$\pm$1.5 minutes) after $t_{III}$ and much shorter durations ($\sim$15-20 minutes). The ion injections start 42$\pm$13 minutes after $t_{III}$ and last for $\sim$170-300 minutes. The inferred solar injection of the 1.67 MeV/nucleon ions has about the same event duration as that of the 0.92 keV electrons, but with a clear delay of $\sim$50 minutes (Figure ~\ref{fig3}(a)), although these ions and electrons have the same velocity ($\sim$0.060c). If ions suffer energy loss due to adiabatic cooling during their propagation in the IPM, they would have traveled faster than that the observed ion energy indicates and thus, their release at the Sun would have been even later \citep{Ruf95,Koc98}. 

For the other nine electron/$^{3}$He-rich SEP events, the triangular fitting is applied to the in situ electron observations, but not to the in situ ion observations (because of poor ion statistics). Instead, the start time and peak time of solar ion injection are obtained from the start time and peak time of ion intensity-time profiles observed at 1 AU, after subtracting the ion travel time along a 1.2 AU path length. The estimated uncertainties of ion injections are dominated by those of the in situ observations, typically on the order of $\pm$1 hour (Figure~\ref{fig3}(b)). Assuming that the ion injection has equal rise and fall times, the injection duration is inferred to be twice the rise time between the start and peak. 

Figure ~\ref{fig3}(b) shows the derived triangular electron injection profiles (black) at the Sun, as well as the estimated start time (red circles) and peak time (red asterisks) of ion injections, for the December 12, 2002 event (\#10). The injection of low-energy electrons starts 11.7$\pm$6.6 minutes before the coronal release of type III burst and lasts for $\sim$60-150 minutes; the injection of high-energy electrons begins 6.8$\pm$1.2 minutes after the type III burst and has short durations ($\sim$10-20 min); the injection of ions begins 61$\pm$62 minutes after the type III burst and lasts for $\sim$220-420 minutes. A clear delay is also evident between the inferred injections of ions and electrons with similar velocities. For the other eight events (\#1-3 and 5-9), the estimated start time (see Figure ~\ref{fig4}(b)) and duration of electron and ion injections are listed in Table~\ref{tbl-1}.

\section{Association with other solar phenomena}

All the selected electron/$^{3}$He-rich SEP events have an associated type III radio burst (Table~\ref{tbl-1}). For eight events (\#2-4 and 6-10), the associated type III burst is observed by WIND/WAVES from in situ local plasma frequency ($\sim$10-30 kHz) up to 14 MHz. Its coronal release time, $t_{III}$, can be estimated from the in situ time of maximum emissions of $\sim$14 MHz observed at 1 AU, after subtracting the photon travel time of 500 s along a 1 AU path (Figure 4(a)). In these eight events, low-energy ($\sim$0.4 to 9 keV) electron injections start 1-30 minutes before $t_{III}$, while high-energy ($\sim$10 to 300 keV) electron injections begin 0-17 minutes after $t_{III}$. For the other two events (\#1 and 5), the associated type III radio burst is observed by WIND/WAVES only up to 1-2 MHz, and thus, $t_{III}$ cannot be accurately obtained (with uncertainties $\gtrsim$ $\pm$10 minutes). 

Among the ten electron/$^{3}$He-rich SEP events, only three (\#2, 5, 8) have a reported GOES SXR burst associated with the inferred start of electron injections and/or ion injections. Only three (\#2, 3, 8) have an associated H$\alpha$ flare, respectively, located at solar longitude W24$^{\circ}$, W54$^{\circ}$ and W63$^{\circ}$, which could be magnetically connected to the vicinity of Earth. Such poor associations with solar flares are typical of electron/$^{3}$He-rich SEP events \citep{Wang12}. However, some of these events may have unreported flares (H. Hudson and N. Nitta, private communication), partially occulted flares \citep{Krk07}, and/or microflares \citep{Chr08, Han08}.

Nine out of the ten events (except \#1) have coverage from the SOHO/LASCO coronagraph and all  nine have an associated CME (Table~\ref{tbl-1}) taking off from the western hemisphere of the Sun that is magnetically well connected to the WIND and ACE spacecrafts. The associated CMEs move at speeds of $\sim$580-1100 km/s, compared to the yearly average CME speed of $\sim$300-500 km/s \citep{Yas04}. Six have an angular width of 7$^{\circ}$-33$^{\circ}$, two are 51$^{\circ}$-62$^{\circ}$ wide, and one is 140$^{\circ}$ wide, compared to the average width of $\sim$47$^{\circ}$-61$^{\circ}$ for normal CMEs (20$^{\circ}$$<$ Width $<$120$^{\circ}$) \citep{Yas04}. Thus, all the associated CMEs are fast and most of them are narrow.

From the LASCO observations of the CME altitude versus time, we can estimate where the CME nose is located at the inferred start of solar particle injections by linear interpolation and extrapolation, for the selected SEP events. Figure~\ref{fig4}(a) shows that for the June 4, 2000 event (event \#4), the associated CME appears to launch (near the photosphere) around the inferred start time of low-energy electron injections, and reaches an attitude of $\sim$0.5 $R_{S}$ at the start of high-energy electron injections and of $\sim$1.7-3.0 $R_{S}$ at the start of ion injections. Figure~\ref{fig4}(c) plots the projected CME altitudes at the start of inferred electron and ion injections for the selected electron/$^{3}$He-rich SEP events. These CMEs are at altitudes of $\sim$0.5-2.0 $R_{S}$ and $\sim$2-11 $R_{S}$, respectively, at the start time of high-energy electron injections and ion injections. Low-energy electron injections appear to begin near the extrapolated CME launch in six events (\#2-4, 6, 7, 10), before the CME launch in two events (\#8, 9) and after in one event (\#5).

Only for event \#8 (Table~\ref{tbl-1}), a coronal type II burst is reported at 25-180MHz (SGD), indicating the presence of a shock wave. At onset of this type II burst, the associated CME has reached an altitude of $\sim$2.8 $R_{S}$, where the local plasma frequency should be much lower than the observed type II frequency range, suggesting that this type II burst may be produced at the CME flanks at lower altitudes. However, different density models could significantly affect the determination of plasma frequency \citep{Leb98}.

\section{Summary and Discussion}

We have examined the timing of electron and ion injections at the Sun for ten good electron/$^{3}$He-rich SEP events, using a 1.2 AU particle path length. On average, the injection of low-energy ($\sim$0.4 to 9 keV) electrons starts first and lasts for $\sim$90 to $\sim$160 minutes, while the injection of high-energy ($\sim$10 to 300 keV) electrons starts 17$\pm$3 minutes after the start of the low-energy electron injection and lasts for only $\sim$10 to $\sim$30 minutes, ending before the low-energy electron injection reaches maximum. The injection of $\sim$0.1-2 MeV/nucleon ions begins 75$\pm$14 minutes after the start of the low-energy electron injection, and lasts for $\sim$200 to $\sim$550 minutes. 

The selected electron/$^{3}$He-rich SEP events have a poor association with H$\alpha$ flares or GOES SXR bursts, but all have a type III burst and a west-limb CME observed by SOHO/LASCO (when its observations are available). On average, the associated CMEs appear to launch near the start of inferred low-energy electron injections, and reach an altitude of 1.0 $R_{S}$ and 4.7 $R_{S}$ at the start of high-energy electron injections and ion injections, respectively.

In these selected SEP events, the electron and ion path lengths that are estimated from the velocity dispersion analysis are consistent with a nominal 1.2 AU length for the Parker spiral field, within the estimated uncertainties ($\sim$0.1-0.2 AU). For electrons, the observed nearly scatter-free temporal profiles and pitch-angle distributions suggest a mean free path $\gtrsim$1 AU \citep{Lin74,Wang11}, indicating a 1.2 AU path length as a good approximation. For ions, the observations exhibit prompt arrivals over a broad energy range ($\sim$0.01-2 MeV/nucleon, see Figure~\ref{fig1}(c) for example). Numerical simulations of ion propagation \citep{Lit04,Sai05,Lai10} suggest that the inferred ion path lengths may have significant errors, but for events with weak scattering in the IPM, the errors in injection times are, typically, only  several minutes, much smaller than the uncertainties in, and the difference between, the injection times estimated in this study. On the other hand, these injection times are inferred by fitting to the observations from the event onset through the peak (that are likely dominated by particles with little/weak scattering), not to the observations in the decay tail (that are probably dominated by scattered/reflected particles). Future investigations of detailed particle propagation will require a focused transport modeling (including the effects of adiabatic cooling, magnetic focusing, and pitch angle scattering) \citep[e.g.,][]{Zha99,Qin06,Li08,Mas12}, which  exceeds the scope of this paper.

The selected ten electron/$^{3}$He-rich events all have a type III burst, and eight events have a clearly determined release of type III burst at the Sun, $t_{III}$. Low-energy electron injections start 1-13 minutes earlier than $t_{III}$ in six events (\#2-4, 6, 7, 10), and 20-30 minutes earlier in the other two events (\#8 and 9) with the in situ rise of low-energy electrons superimposed on the decay of a previous event (see Figure~\ref{fig1}(a)). These suggest that low-energy electrons are responsible for the generation of type III radio emissions, which is consistent with previous studies of impulsive electron injections \citep{Wang06b} and in situ Langmuir waves \citep[e.g.,][]{Erg98}. Based on simulations, \citet{Kon09} suggest that solar impulsive electrons are injected simultaneously at all energies at the Sun, but the low-energy electrons are significantly decelerated by beam-plasma interactions in the IPM, to produce the inferred earlier injection before high-energy electrons. However, this scenario cannot explain the delayed injection of high-energy electrons after the release of type III bursts at the Sun.  

If the accelerated electrons and ions are temporarily stored (e.g., in closed coronal loops) before their release into the IPM, e.g., caused by the passage of a CME, then the inferred solar particle injections should exhibit no delay. In all the selected events, however, the inferred solar ion injections show a clear delay after the electron injections, even for ions and electrons with the same velocity. If solar energetic ions suffer energy loss due to adiabatic cooling during propagation in the IPM \citep{Ruf95, Koc98}, they would have traveled faster than that the observed ion energies indicate, and their solar release would become even later. Hence, in these events, the derived particle injections likely reflect the solar acceleration of low-energy electrons, of high-energy electrons, and of ions that occur at different times.

In these electron/$^{3}$He-rich events, the injections of high-energy electrons show clearly delayed starts but much shorter durations, compared to the injections of low-energy electrons. Such injection delay cannot be due to scattering in the IPM. In addition, the observed electron energy spectra all show a single smooth power law across the transition between low-energy and high-energy electrons (see Figure~\ref{fig2}(f) for example). These results suggest that the low-energy electron injections may provide seed electrons for the delayed acceleration to high energies \citep{Wang06b}, which is probably related to a propagating CME.

For the ten selected events, their 100\% association with fast, mostly narrow, west-limb CMEs is remarkable since only $\sim$30\% of $^{3}$He-rich electron events are associated with fast, narrow CMEs (although $\sim$60\% have a west-limb CME) \citep{Wang12}. Our selection criteria favor intense ion events, suggesting that the strong acceleration of $^{3}$He-rich ions may be associated with fast, narrow CMEs. On average, the inferred ion injections start when the associated CMEs reach an altitude of 4.7 $R_{S}$. However, to produce the high ionization states that increase with energy \citep[see][for events \#1-4 in this study]{DiF08}, which are typically observed in impulsive SEP events \citep{Pop06}, requires a $N\tau_{A}$ $\sim$$10^{10}$ - $10^{11}$ s$\cdot$cm$^{-3}$, where $N$ is the plasma density and $\tau_{A}$ is the acceleration timescale \citep[e.g.,][]{Koc00, Kat06}. Since $\tau_{A}$ should be less than the derived ion injection duration (as short as 200 minutes), $N$ must be $\gtrsim$$10^{6}$ - $10^{7}$ cm$^{-3}$, which corresponds to source altitudes of $\lesssim$ 0.5 $R_{S}$ (depending on the density model). This suggests that the ion acceleration may occur at the lower-altitude flanks (not at the nose) of CMEs.

The ion initial acceleration could be due to the resonance with wave modes that can preferentially heat/accelerate $^{3}$He and heavy ions due to their charge-to-mass ratios \citep[e.g.,][]{Fis78}. \citet{Tem92} and \citet{Rot97} noted that in the terrestrial aurora, precipitating electron beams generate oblique electromagnetic ion cyclotron waves that resonate with the gyrofrequency of $^{3}$He or the second harmonic of the gyrofrequencies of heavy ions, to accelerate these ions. They suggested that a similar process might occur at the Sun in electron/$^{3}$He-rich SEP events. Other studies also propose the resonant acceleration of ions by other wave modes \citep[e.g.,][]{Mil93,Mil98, Liu06, Pae03}. 

\begin{acknowledgements}
This research at Peking University is supported in part by NSFC under contracts 41421003, 41274172, 41474148, 41231069, 41222032, 41174148, and 41474147. The work at UAH is supported by NASA grants NNX14AC08G and NNX15AJ93G. G. Mason is supported by by NASA grant NNX13AR20G.
\end{acknowledgements}



\onecolumn

\begin{landscape}
\tiny
\setlength\LTleft{0pt}
\setlength\LTright{0pt}
\setlength{\tabcolsep}{1.2mm}
\begin{longtable}{cccccccccccccccccc}
\caption{the Ten Electron/$^{3}$He-rich SEP Events\label{tbl-1}}\\
\hline\hline
 N & Date & \multicolumn{3}{c}{Ions} &  &
\multicolumn{2}{c}{Electrons below $\sim$6-9 keV} &  &
\multicolumn{2}{c}{Electrons above $\sim$10 keV}  & Type III/II & 
GOES SXR\tablefootmark{c} & H$\alpha$ Flare\tablefootmark{c}  & 
\multicolumn{4}{c}{CME\tablefootmark{d}} \\
\cline{3-5} \cline{7-8} \cline{10-11} \cline{15-18} \\
 &   & $^{3}$He/$^{4}$He\tablefootmark{a} & 
Release\tablefootmark{b} & Duration\tablefootmark{b}  &   
  & Release\tablefootmark{b}  & Duration\tablefootmark{b}    & 
 & Release\tablefootmark{b} & Duration\tablefootmark{b} & 
Release  &  Class & Location & Launch\tablefootmark{e}  & 
V  & P.A.  & Width \\
 &   &    & (UT) & (min)  &   & (UT)  & (min)  &  & (UT) &  
(min) & (UT)  &   &  & (UT)   & (km/s)  & (deg)  & (deg)  \\
(1) & (2)   & (3)  & (4) &  (5)  &  & (6)   & (7)  & & (8) &  
(9) & (10) &  (11)  & (12) & (13)  & (14)  & (15)  & (16) \\ 
\hline
1 & 1998 Aug 18 & 1.32$\pm$0.26 & 02:41$\pm$00:31 & 190-320 & & 
01:58$\pm$00:08 & 90-170 & & 02:12$\pm$00:04.2 & 12-60 & III & & & 
… & … & … & …\\ 
2 & 1999 Aug 7 & 1.52$\pm$0.08 & 18:37$\pm$00:50 & 280-530 && 
16:45$\pm$00:06 & 50-130 & & 17:02$\pm$00:01.2 & 8-12 & III/16:55 & 
M1.2 & S15W24 & 16:42 & 577 & 297 & 7 \\
3 & 2000 Apr 1 & 0.11$\pm$0.02 & 19:53$\pm$00:37 & 200-530 && 
19:04$\pm$00:07 & 60-230 & & 19:20$\pm$00:03.9 & 20-25 & 
III/19:05 & &N10W54 & 18:50 & 586 & 291 & 33 \\
4 & 2000 Jun 4 & 0.30$\pm$0.02 & 07:36$\pm$00:13 & 170-300 &&
06:41$\pm$00:09 & 80-150 & & 07:01$\pm$00:01.5 & 15-23 &
III/06:54 & & & 06:51 & 597 & 295 & 17 \\
5 & 2002 Jun 30 & 0.16$\pm$0.06 & 11:54$\pm$00:43 & 290-800 && 
08:58$\pm$00:12 & 130-270 & & 09:06$\pm$00:01.9 & 20-30 &
III & C2.1 & & 08:28 & 623 & 299 & 30 \\
6 & 2002 Sep 22 & 0.54$\pm$0.02 & 24:04$\pm$00:29 & 170-250 && 
23:32$\pm$00:04 & 80-120 & & 23:36$\pm$00:02.0 & 15-40 &
III/23:36 & & & 23:31 & 960 & 287 & 140 \\
7 & 2002 Sep 24 & 0.75$\pm$0.05 & 11:50$\pm$00:25 & 170-280 && 
10:52$\pm$00:06 & 100-180 & & 11:07$\pm$00:01.5 & 8-18 &
III/10:50 & & & 10:56 & 1104 & 277 & 62 \\
8 & 2002 Oct 20 & 1.47$\pm$0.05 & 14:46$\pm$00:33 & 240-570 && 
13:33$\pm$00:07 & 70-130 & & 14:06$\pm$00:01.8 & 5-40 &
III/14:03 & C6.6 & S13W63 & 13:52 & 1011 & 247 & 20 \\
 &  &  &  &  & &  & & & & & II/14:24 &  & & &  &  &  \\
9 & 2002 Dec 12 & 1.11$\pm$0.07 & 01:24$\pm$00:43 & 190-400 && 
00:11$\pm$00:05 & 80-130 & & 00:35$\pm$00:01.4 & 17-18 &
III/00:32 & & & 00:33 & 624 & 297 & 13 \\
10 & 2002 Dec 12 & 0.39$\pm$0.03 & 13:29$\pm$01:02 & 220-420 && 
12:16$\pm$00:07 & 60-150 & & 12:35$\pm$00:01.2 & 10-20 &
III/12:28 & & & 12:13 & 723 & 287 & 51 \\

\hline
\end{longtable}
\normalsize

\tablefoot{\\
$^{a}$The $^{3}$He/$^{4}$He ratio is calculated at 0.5-2.0 MeV/nucleon.\\
$^{b}$The release is energy-averaged;  the duration is energy-dependent.\\
$^{c}$The flare information is from the Solar Geophysical Data (SGD).\\
$^{d}$The CME information is from the SOHO LASCO CME catalog (http://cdaw.gsfc.nasa.gov/CME\_list/).\\
$^{e}$The CME launch time is estimated from linear extrapolation to the photosphere, after subtracting the photon travel time of 500 seconds.
}

\end{landscape}

\end{document}